%% file: ITER_wall_forces.tex
\documentclass[twoside,twocolumn]{article}
\usepackage{hhline}
\usepackage{authblk}
\usepackage{multicol}

\input{./packages.tex}

\def\VecA{\mathbf{A}}
\def\VecV{\mathbf{v}}

\def\VecB{\mathbf{B}}

\def\VecJ{\mathbf{J}}

\def\TensD{\underline{\pmb{\text{D}}}}
\def\TensV{\underline{\pmb{\tau}}}
\def\TensK{\underline{\pmb{\kappa}}}

\newcommand{\pderiv}[2]{\frac{\partial #1}{\partial #2}}       

\begin{document}

\title{Complete 3D MHD simulations of the current quench phase of ITER mitigated disruptions}

\renewcommand\Affilfont{\itshape\small}
\author[1,2]{F.J. Artola}
\author[2]{A. Loarte}
\author[1]{M. Hoelzl}
\author[2]{M. Lehnen}
\author[1]{N. Schwarz}
\author[3]{the JOREK team}
\affil[1]{Max Planck Institute for Plasmaphysics, Boltzmannstr. 2, 85748 Garching, Germany}
\affil[2]{ITER Organization, Route de Vinon sur Verdon, 13067 St Paul Lez Durance Cedex, France}
\affil[3]{please refer to [M Hoelzl, G T A Huijsmans, S J P Pamela, M Becoulet, E Nardon, F J Artola, B Nkonga et al, Nuclear Fusion 61, 065001 (2021)].}

\date{}
\twocolumn[
  \begin{@twocolumnfalse}
    \maketitle
    \begin{abstract}
Complete 3D simulations of the current quench phase of ITER disruptions are key to predict asymmetric forces acting into the ITER wall. We present for the first time such simulations for ITER mitigated disruptions at realistic Lundquist numbers. For these strongly mitigated disruptions, we find that the safety factor remains above 2 and the maximal integral horizontal forces remain below 1 MN. The maximal integral vertical force is found to be 13 MN and arises in a time scale given by the resistive wall time  as expected from theoretical considerations. In this respect, the vertical force arises after the plasma current has completely decayed, showing the importance of continuing the simulations also in the absence of plasma current. We conclude that the horizontal wall force rotation is not a concern for these strongly mitigated disruptions in ITER, since when the wall forces form, there are no remaining sources of rotation.
    \end{abstract}

\medskip
    
\hspace{0.6 cm} Corresponding author: \textbf{F.J. Artola}
\hspace{2.5cm} Email: \textcolor{blue}{javier.artola.such@ipp.mpg.de}
\vspace{1cm}
  \end{@twocolumnfalse}
]

\section{Introduction}
There is no general consensus on the predictions for the maximum integral wall forces acting on the vacuum vessel during ITER disruptions.  Although it is broadly accepted that the maximum total vertical force ($F_z$) will be of the order of 80 MN \cite{Sugihara2007_forces,Clauser_2019}, the prediction for the horizontal force ($F_h$) differs by orders of magnitude. Using the maximum values found in JET for $F_h$ ($\sim 4$ MN) \cite{gerasimov2015jet} and extrapolating with  Noll's formula \cite{noll1997present}, maximal horizontal forces of 40 MN are obtained. Dedicated studies using a source/sink model inspired by JET measurements \cite{Riccardo_2000} also find $F_h\sim 40 $MN \cite{bachmann2011specification}. However, other models based on a 1/1 kink mode that interacts with the vacuum vessel only via eddy currents, produce forces an order of magnitude smaller \cite{Pusto2021}. The ratio between the vessel's current decay time ($\tau_w$) and the current quench time ($\tau_{CQ}$) was found to play a major role determining the horizontal force  with 3D MHD simulations \cite{strauss2018reduction}. In that reference, the maximum force for ITER was $F_h\sim 30$ MN when $\tau_{CQ}/\tau_w \gg 1$ and the minimum force was $F_h\sim 4$ MN when $\tau_{CQ}/\tau_w \ll 1$. Two main effects could be the cause of the strong dependence of the vessel forces with $\tau_{CQ}$. In the following we will refer to the ITER vacuum vessel as "wall" for simplicity.

\medskip

 The first effect is related to the evolution of the edge safety factor ($q_a$). If $q_a$ remains above 2 during a disruption, the 1/1 mode leading to large horizontal forces is not expected. Whether  $q_a$ remains above 2 during a disruption depends on the competition between the plasma current decay and the vertical motion of the plasma column (i.e. plasma volume shrinkage). If the vertical motion takes place at constant plasma current (i.e. $\tau_{CQ}/\tau_w \gg 1$), it is expected that $q_a$ decreases below unity \cite{artola2020understanding} and that 1/1 modes become unstable. On the other hand, if the plasma current ($I_p$) decay is faster than the vertical motion,  $q_a$ increases over time since $q_a\propto 1/I_p$. However, even in the limit where $\tau_{CQ}/\tau_w \ll 1$, elongated plasmas drift vertically in a time scale given by $\tau_{CQ}$ \cite{Kiramov_2018,artolasuch:tel-02012234}. Therefore, the evolution of $q_a$ will be ultimately determined by the function that describes the vertical position as a function of the plasma current ($Z(I_p)$), which is dictated by the geometry of the plasma and the wall in this fast current quench limit.

\medskip

The second effect arises from the penetration time of the magnetic field across the wall ($\tau_w$). It was derived in \cite{Pustovitov_2017}, that the total wall force can be computed with a surface integral enclosing the wall and plasma volumes
\begin{equation}
\label{eq:force}
\v{F} \equiv \frac{1}{\mu_0}\int_{wall} \v{J}\times\v{B} dV =  \int_{wall+}  \left( (\v{B}\cdot\v{n})\v{B} - \frac{B^2}{2}\v{n} \right) dS
\end{equation}
where $\v{F}\equiv(F_x, F_y, F_z)$, $\v{J}$ is the current density, $\v{B}$ is the total magnetic field and $\v{n}$ is a unit vector perpendicular to the toroidal surface. Therefore, for time scales much shorter than $\tau_w$, the vessel forces remain small ($\v{F}\approx 0$) since $\v{B}$ remains approximately unchanged outside the wall. However, as it is deduced from \eqref{eq:force} and it is shown in this paper, the force can arise after the field penetrates the wall even in the absence of plasma currents (this was also observed in \cite{Vadim2021}). This fact has been largely ignored in the literature of 3D MHD disruptions and 3D simulations are typically not continued after the plasma current vanishes. In the present work we make sure to extend the simulation time beyond $\tau_w$ to allow $\v{B}$ to penetrate the vacuum vessel.

\medskip

A major finding of our studies is that, in the $\tau_{CQ}/\tau_w \ll 1$ limit, wall forces are static in the toroidal direction (i.e. non-rotating). This is due to the fact that when the wall forces are maximum (when the wall currents decay inside the vacuum vessel) the plasma current has already vanished and thus there is no electromagnetic source to drive the rotation. Therefore the issue of resonances between the force rotation and the natural frequencies of the vacuum vessel \cite{schioler2011dynamic} is not relevant for this limit. To our knowledge this fact has  not been identified in the literature so far. 

\medskip

The mitigation of electromagnetic loads in ITER \cite{lehnen2015disruptions} relies on the dissipation of magnetic energy from the plasma by electromagnetic radiation together with the experimental evidence that the wall forces decrease at smaller $\tau_{CQ}$ \cite{Pautasson2016_reduction}. Consequently, the ITER disruption mitigation system aims to reduce the plasma temperature during the CQ and thereby $\tau_{CQ}$ by using massive material injection. In this paper, we demonstrate with 3D MHD simulations featuring realistic time scales, that  the wall integral forces are indeed largely reduced for mitigated disruptive plasma conditions with respect to the maximal values extrapolated from simple models and empirical assumptions described above.

\medskip

In section \ref{sec:model}, we present the MHD model and in section \ref{sec:initial}, we present the simulation setup (initial conditions and parameters). Finally, in section \ref{sec:results}, we present our results and we summarize our conclusions in section \ref{sec:conclusions}.

\begin{figure}[h]
  \centering
  \includegraphics[width=0.5\textwidth]{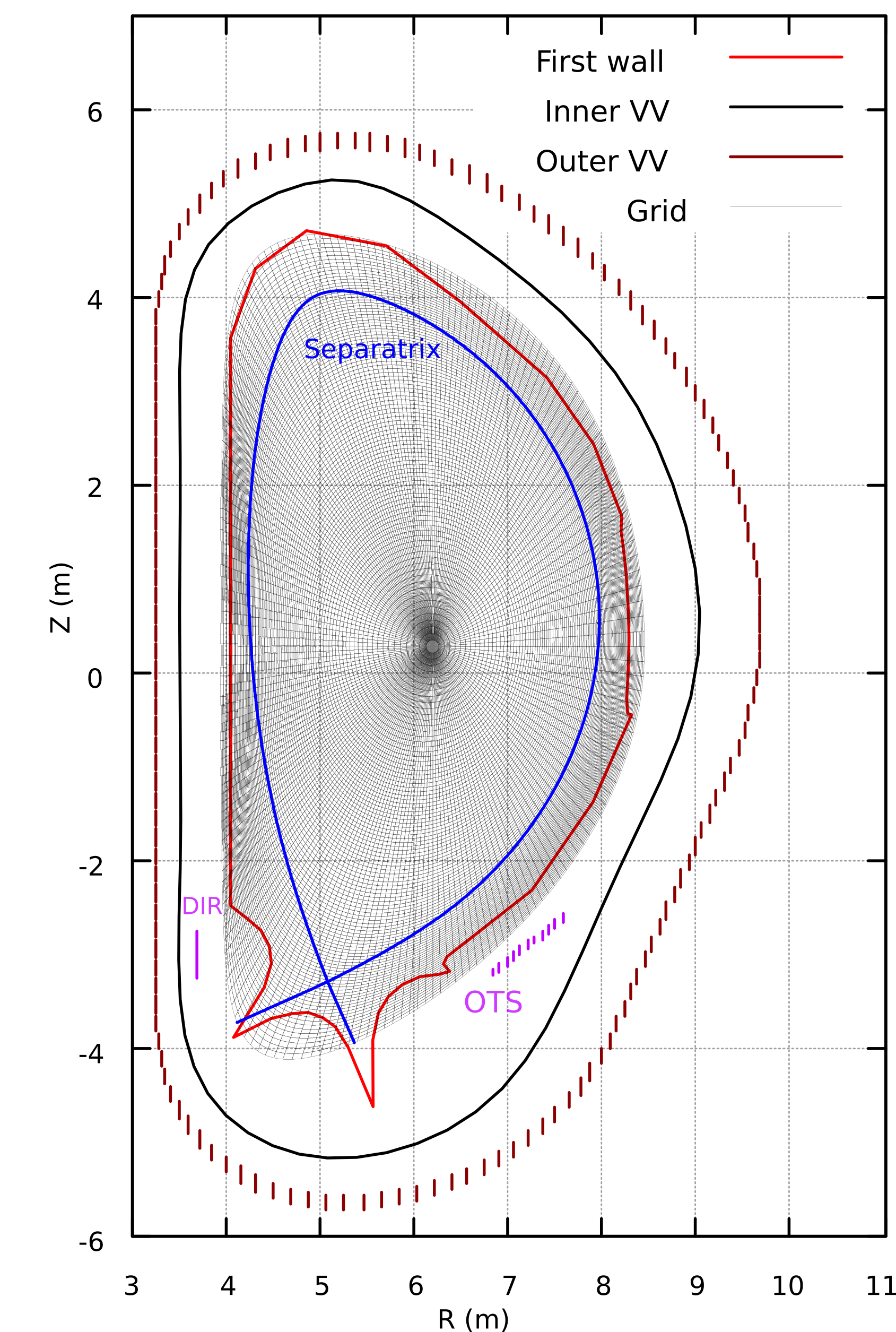}
\caption{Computational grid and passive components (OTS, DIR, inner and outer vessel) included in our simulations. The ITER first wall (red)  and the initial plasma separatrix (blue) are included for reference. Except for the inner vacuum vessel layer, where 3D currents can flow, the current path in the other passive components is toroidal and axisymmetric since they are discretized by set of toroidal filaments.}
\label{fig:grid_wall}
\end{figure}

\begin{figure}[h]
  \centering
  \includegraphics[width=0.48\textwidth]{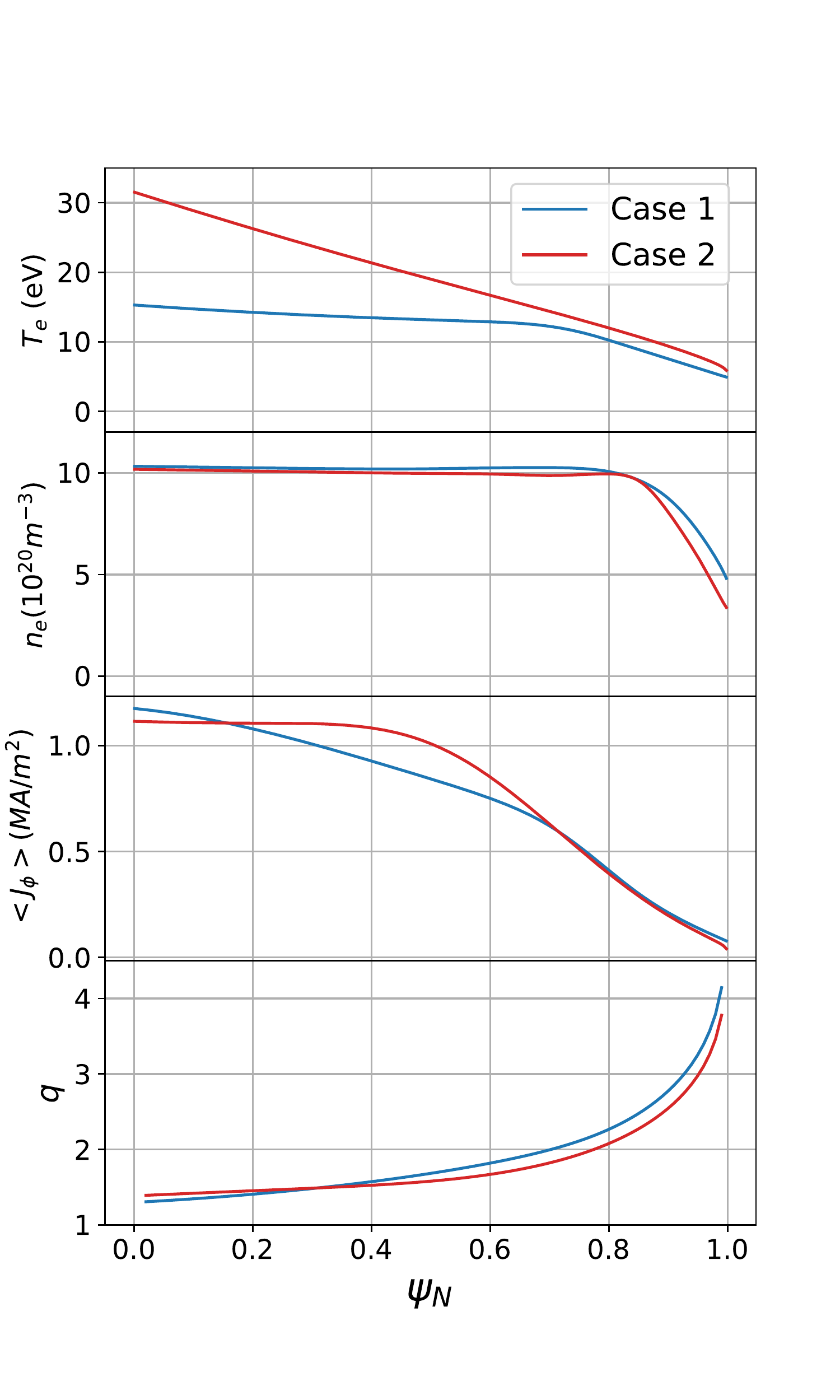}
\caption{Profiles at the start of the 3D simulations for the two cases considered in this paper. From top to bottom the profiles of the electron temperature, electron density, poloidally averaged toroidal current profile and safety factor profiles are shown. The radial coordinate is the normalized poloidal magnetic flux. }
\label{fig:eq_profiles}
\end{figure}

\section{MHD model and assumptions} 
\label{sec:model}

The basic model employed in this paper is a single temperature visco-resistive MHD model \cite{Hoelzl_2021}

\begin{align}
\pderiv{\VecA}{t} &= \VecV\times\VecB - \eta\VecJ  - \nabla \Phi, \label{eq:mhd:A}\\
\rho\pderiv{\VecV}{t} &= -\rho\VecV\cdot\nabla\VecV - \nabla p +
\VecJ\times\VecB + \nabla\cdot\TensV,
\label{eq:mhd:v} \\
\pderiv{\rho}{t} &=
-\nabla\cdot(\rho\,\VecV)
+\nabla\cdot(\TensD\nabla\rho)
,\label{eq:mhd:rho}
\\
\pderiv{p}{t} &= -\VecV\cdot\nabla p
- \gamma p\nabla\cdot\VecV
+ \nabla\cdot(\TensK\nabla T) 
\label{eq:mhd:p}
\end{align}
that uses the following reduced MHD ansatz for the plasma flow ($\VecV$) and the magnetic field ($\VecB$)
\begin{align}%
\VecB &= \nabla\psi\times\nabla\phi + F_0 \nabla\phi, \label{eq:B}\\
\VecV &= -\frac{R^2}{F_0}\nabla\Phi\times\nabla\phi +\VecV_\parallel, \label{eq:v}
\end{align}
where $\psi$ is the poloidal magnetic flux and $F_0 = RB_\phi$ is taken as a constant.
Poloidal currents evolve according to  current conservation and momentum balance, but we neglect their contribution to the toroidal field \cite{Hoelzl_2021}. Successful axisymmetric and 3-dimensional benchmarks of VDEs have been performed with the full MHD codes NIMROD and M3D-C$^{1}$ to check the validity of these assumptions for the wall forces \cite{krebs2020axisymmetric,artola3DVDE_bench}.  The quantities shown in equations (2)-(5) are  the magnetic vector potential ($\VecA$), the ion density ($\rho$), the total pressure ($p$), the total temperature ($T\equiv T_e + T_i$), the electrostatic potential ($\Phi$) and the current density ($\VecJ$). Other parameters in equations (2)-(5) are the plasma resistivity ($\eta$), the stress tensor($\TensV$), the thermal  conductivity ($\TensK$) and the particle diffusion coefficients ($\TensD$) and the ratio of specific heats ($\gamma$). The thermal conductivity coefficient tensor $\TensK$ presents a high anisotropy (i.e. $\kappa_\parallel \gg \kappa_\perp$) while the particle diffusion coefficients are normally isotropic. 

\medskip

The equations are numerically solved with the fully implicit JOREK code\cite{Hoelzl_2021, huysmans2007mhd}. Fourier harmonics are used to represent the toroidal direction and for the poloidal plane, third order quadrilateral Bezier finite elements are used (see figure \ref{fig:grid_wall}). To take into account resistive wall effects, JOREK is implicitly coupled to the STARWALL code \cite{Merkel2015,Hoelzl2012a,artolasuch:tel-02012234}.  

\medskip

Our modeling for ITER mitigated disruption simulations does not consider either Ohmic heating or radiated power by impurities during the current quench. By neglecting them, we assume that their associated terms that should appear in equation \eqref{eq:mhd:p} exactly cancel each other. In other words, we assume that the plasma magnetic energy is completely radiated. Such an assumption is realistic since radiation efficiencies $ > 90 \%$ are required to mitigate disruptions in ITER DT plasmas \cite{lehnen2015disruptions}. In our case, there are no sources of temperature and density, therefore they are determined by the initial conditions and their evolution through the convection and diffusion/conduction terms.

\medskip

We model the ITER vacuum vessel composed by  inner layer and outer axisymmetric layers using the thin wall approximation (see figure \ref{fig:grid_wall}). The resistivity and thickness of each layer is given according to the vessel design specifications ($\eta_w=0.8 \, \mu\Omega$ and $d_w =6$ cm). 3D currents are allowed to flow in the inner layer, which is discretized with  75000 thin linear triangles. The outer layer is discretized with toroidal filaments and therefore only axisymmetric toroidal currents flow there. Since we simulate current quench times that are faster than the field penetration time of the inner vessel layer, 3D currents in the outer vessel do not play a role for the plasma dynamics in our case. The outer triangular support (OTS) and the divertor inboard rail (DIR) are also included since they are important passive components for vertical stability. This model for the ITER passive structures has already been benchmarked successfully with the code DINA for axisymmetric simulations \cite{artola2018non,overview}. Note that we are not taking into account the ITER blanket modules, which could potentially lead to additional stabilizing effects since dipolar currents with characteristic time scales of $\sim$10 ms can be induced in these modules. However, toroidal currents cannot circulate directly from module to module since they are insulated from each other and only electrically connected through the vacuum vessel. For these reasons, we do not expect that the presence of blanket modules changes the results presented in this work significantly (i.e. the maximum wall forces).

\medskip

 The computational polar grid (see figure \ref{fig:grid_wall}) is composed of $100\times 200$ radial and poloidal Bezier elements, respectively. The plasma computational boundary roughly matches the ITER first wall but does not represent the details of the divertor structures adequately. More refined boundaries including sharp edges will be included in future work.

\section{Initial conditions and used parameters} 
\label{sec:initial}

\begin{table*}
\small
\centering
\def\arraystretch{1.5}
\begin{tabular}{|c|c|c|}

\hline
\textbf{Parameter} & \textbf{Value}   &  \textbf{Description}\\ \hhline{|=|=|=|}
$ D$ & $  2 \textrm{ m}^2/\textrm{s}$ & Isotropic particle diffusion coefficient \\ \hline
$\kappa_\perp$ & $  2\times 10^{21} \textrm{ (m s)}^{-1}$ & Perpendicular thermal conductivity \\ \hline
$\kappa_\parallel = \kappa_0 (T_e/T_{e0})^{5/2} $ & $ \kappa_0 = 1.77\times10^{28}  \textrm{ (m s)}^{-1}$ & Parallel thermal conductivity  \\ \hline
$\eta = \eta_0 (T_e/T_{e0})^{-3/2} $ & $ \eta_0 = 6.74\times10^{-7} \Omega\textrm{ m}$ & Parallel resistivity \\ \hline
$(\mu_\parallel^*, \mu_\perp)$ & $ (1630, 4.24) \times 10^{-7} \textrm{ kg}/\textrm{(m s)}$ & 
Parallel/perpendicular dynamic viscosity \\ \hline

\end{tabular}%

\caption{Parameters used during the current quench phase. For the temperature dependent parameters the reference temperature is $T_{e0}= 300 $ eV. Note that except for $\eta$ and $\kappa_\parallel$ the coefficients are spatially constant. $^*$Due to numerical problems, $\mu_\parallel$ had to be increased at some points of the simulation by a factor of 10 (see section \ref{sec:halos}). }
\label{tab:param}
\end{table*}

The chosen initial conditions correspond to those of a strongly mitigated disruptive plasma in which the thermal quench (TQ) has already taken place. A 15 MA / 5.3 T L-mode plasma was considered and the initial conditions were constructed with the following procedure. In the first place, the pre-TQ L-mode reference equilibrium was computed. Secondly, in order to achieve a large electron density, the density profile was re-scaled by a factor of 20 while the plasma temperature was re-scaled by a factor of 1/20. This procedure allows to keep the pressure profile unchanged and to obtain an identical Grad-Shafranov equilibrium as for the reference L-mode but with higher electron densities and lower temperatures. Finally, the perpendicular thermal transport coefficient ($\kappa_\perp$) was increased in order to simulate an artificial thermal quench, leading to the profiles shown in figure \ref{fig:eq_profiles} which are used as starting points for the 3D simulations. By adjusting $\kappa_\perp$, the two different cases shown in figure \ref{fig:eq_profiles} were constructed. These two cases are used to study the sensitivity of the current quench dynamics to differences in the temperature and current density profiles.  The resulting temperatures and densities can be expected after the thermal quench following injection of Neon via shattered pellet injection (SPI). Note that the chosen temperatures are close to the Neon radiation peak ($\sim 30 $ eV). Once the profiles shown in figure \ref{fig:eq_profiles} are established, the thermal diffusion coefficients are reduced to avoid a further temperature decay and the 3D simulation of the CQ starts. During the CQ, 11 toroidal Fourier harmonics are considered ranging from 0 to 10 ($n\in [0,10]$). The plasma parameters used during the current quench phase are listed in table \ref{tab:param}. 

\medskip
The safety factor ($q$) profiles are not completely flattened inside the $q=2$ surface, which are typically observed to be the relaxed MHD states after the thermal quench \cite{Nardon2021}. These MHD unstable profiles are chosen intentionally to attain a self-consistent relaxation of the $q$ and $J_\phi$ profiles during the first phase of the CQ. Rather than starting from  arbitrarily stable current density profiles, we consider that in order to obtain a $J_\phi$ profile similar to the one that would be obtained after a TQ, it is more convenient to start with unstable profiles and allow for a self-consistent relaxation.


\medskip

We apply Dirichlet boundary conditions for the fluid variables at the plasma-wall interface ($\rho$, $T$ and $\v{v}$). The temperature and density B.C.s are simply $n_e(t) = 10^{20} \, \textrm{m}^{-3} $ and $T_e(t) = 1$ eV. In addition we set a no-normal flow boundary condition at the plasma-wall interface ($\VecV\cdot\mathbf{n}=0$) and no parallel velocity ($v_\parallel=0$). Note that in our formulation, the $\VecV\cdot\mathbf{n}=0$ condition is equivalent of that of an  ideal wall in the poloidal direction ($\Phi=0$). It also implies that energy and particles cannot lost through the boundary by convection. 

\begin{figure}[h]
  \centering
  \includegraphics[width=0.9\linewidth]{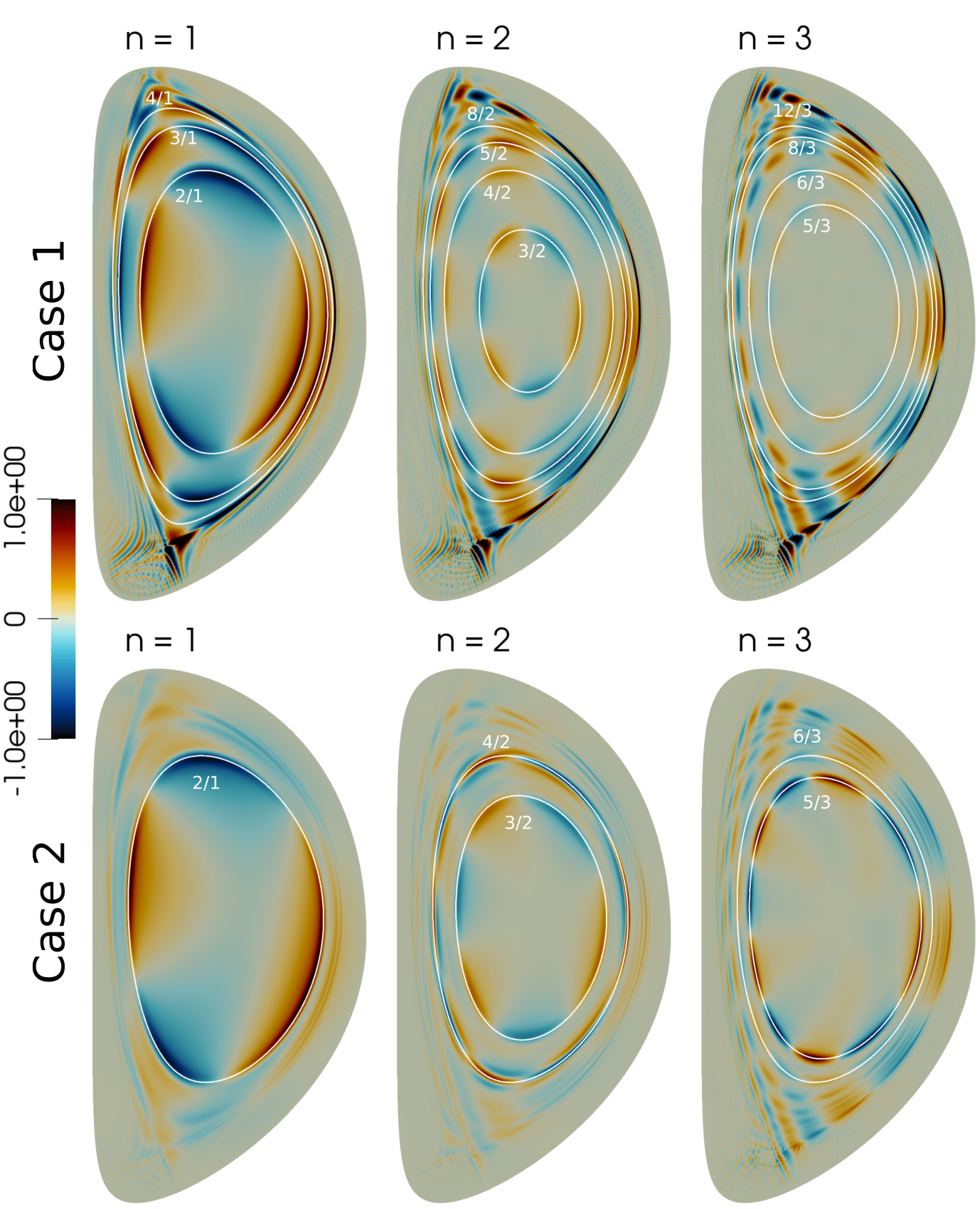}
\caption{Mode structures of the electric potential ($\Phi$) for case \#1 at $t=8.1$ ms and for case \#2 at $t=4.7$ ms. The white contours represent rational flux surfaces $q=m/n$ which are labeled with their associated  poloidal ($m$) and toroidal ($n$) mode numbers. The electric potential is normalized to its maximum value for each mode number.}
\label{fig:modes}
\end{figure}

\begin{figure}[!h]
  \centering
  \includegraphics[width=0.45\textwidth]{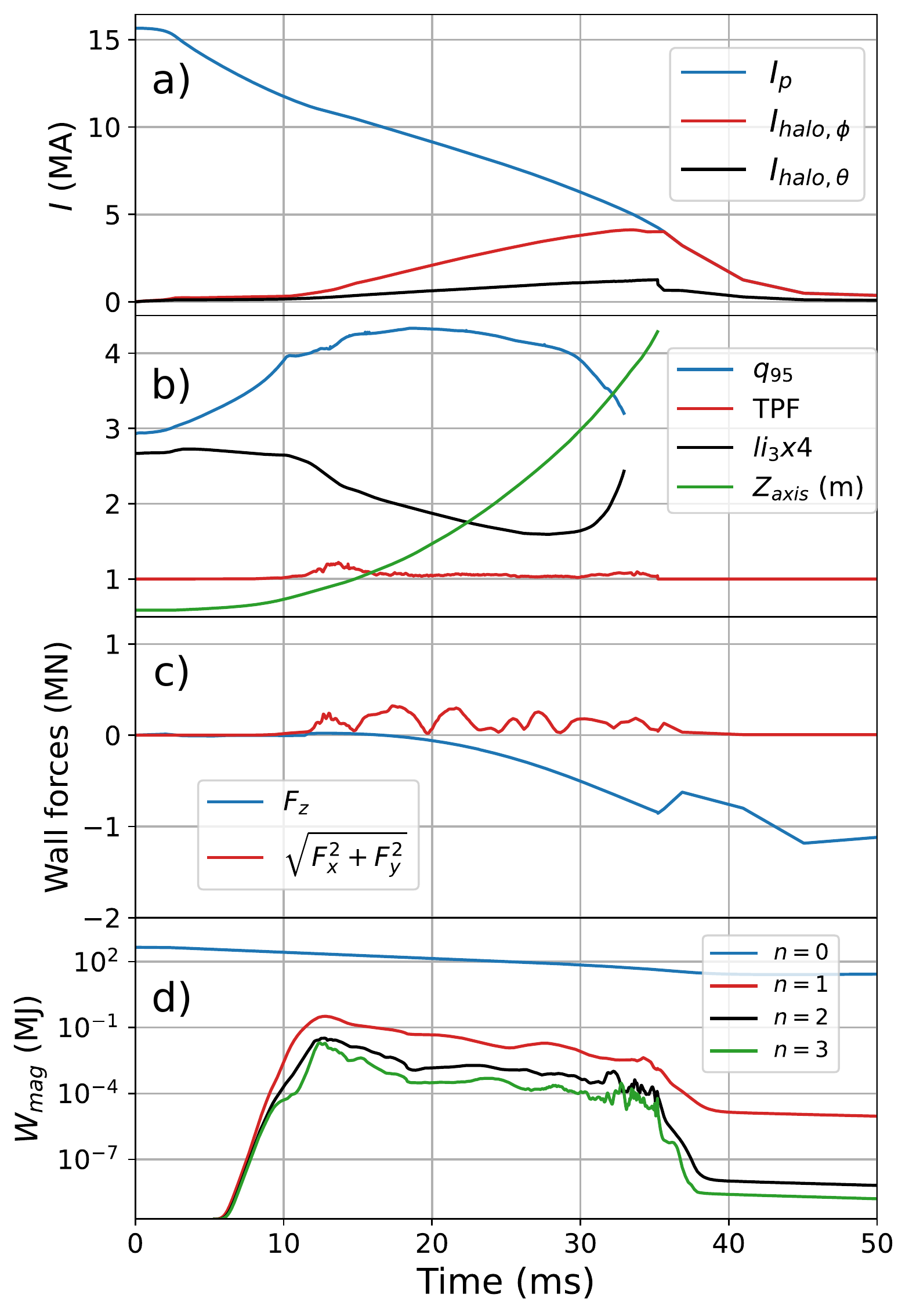}
\caption{Time traces of the CQ simulation for case \#1. (a) Total plasma toroidal current ($I_p$), total poloidal ($I_{halo,\theta}$) and toroidal  ($I_{halo,\phi}$)  halo currents . (b) Edge safety factor ($q_{95}$), toroidal peaking factor of the poloidal halo currents (TPF), internal inductance ($l_i(3)$) and vertical position of the magnetic axis ($Z_{axis}$). (c) Vertical ($F_z$) and horizontal wall forces  ($\sqrt{F_x^2+F_y^2}$). (d) The poloidal magnetic energy of the most dominant toroidal mode numbers ($n=0,1,2,3$).  }
\label{fig:traces_3D_1}
\end{figure}

\begin{figure*}[!h]
  \centering
  \includegraphics[width=0.95\linewidth]{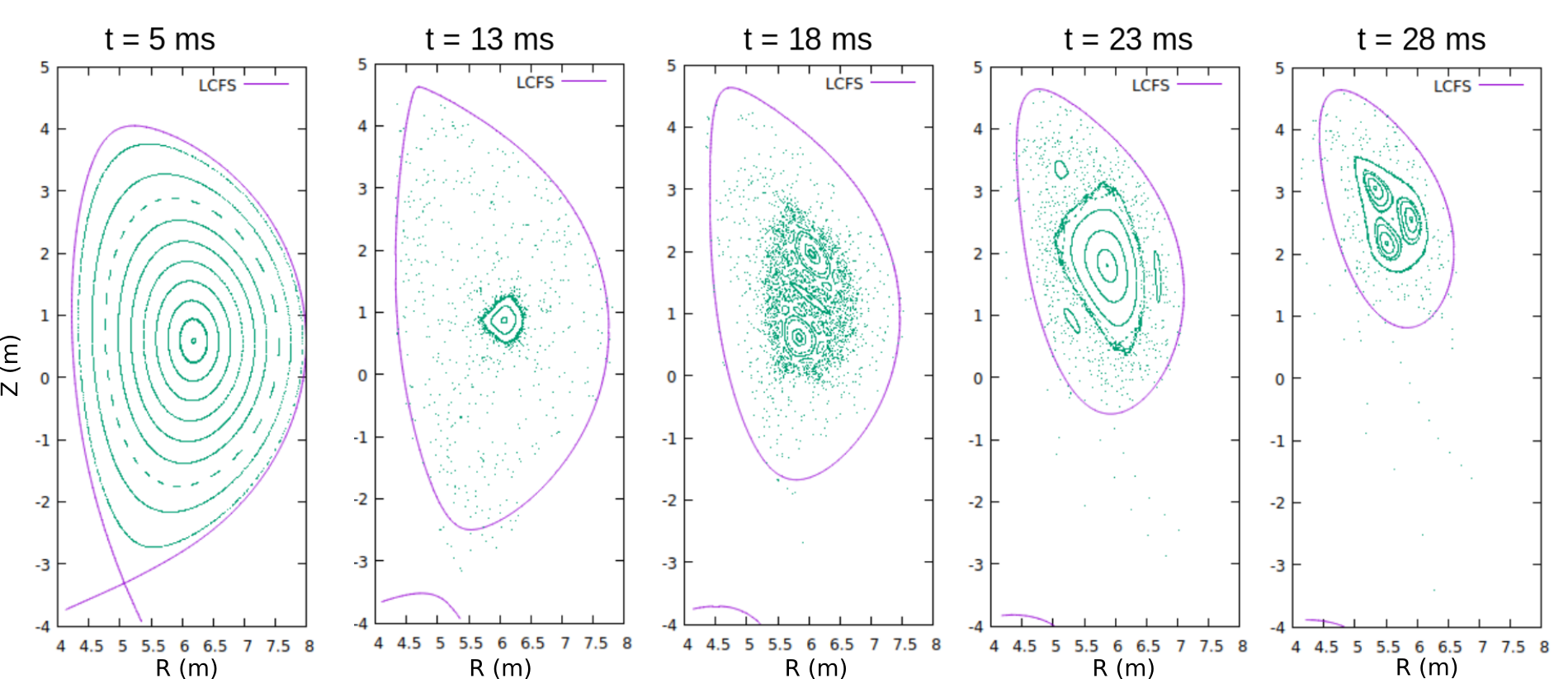}
\caption{Poincare plots for case \#1. The purple contour represent the Last Closed Flux Surface (LCFS).}
\label{fig:poinc_1}
\end{figure*}

\section{ITER current quench simulations} 
\label{sec:results}

In this section we describe and compare the two simulated cases. The main difference between the two cases are the different profiles shown in figure \ref{fig:eq_profiles}, which took $t=5$ ms and $t=2.1$ ms to establish for case \#1 and \#2 by running JOREK axisymmetrically, respectively. Otherwise the employed parameters are the identical (see table \ref{tab:param}).

\subsection{CQ phase of case \#1}
For case \#1, the axisymmetric run lasted about 5 ms to establish the profiles shown in figure \ref{fig:eq_profiles}. At $t=5$ ms the 10 non-axisymmetric modes were initialized to "noise" level and the 3D simulation started. The case is unstable to a variety of tearing modes due to the large plasma resistivity and the initial current profile. These modes are linearly unstable, have similar growth rates ($\sim 2-3$ ms$^{-1}$) and appear at different rational surfaces ($q=m/n$) described by the toroidal ($n$) and poloidal periodicities ($m$) as shown in figure \ref{fig:modes}. The most unstable modes have low$-n$ toroidal mode numbers and the higher$-n$ modes remain subdominant. The evolution of the poloidal magnetic energy for a selection of  dominant mode numbers ($n=1,2,3$) is shown in figure \ref{fig:traces_3D_1}. 

\medskip

At $t=10$ ms, the modes energies start to saturate and reach their maximum value at $t=13$ ms. Note that at the time of mode saturation, the total plasma current has already decayed to 11-12 MA due to the low plasma temperature. Larger "noise" levels for the initial perturbations could cause the saturation phase to occur at a higher $I_p$. In this respect, more realistic initial conditions for the non-axisymmetric modes could be achieved by also simulating the thermal quench, however this is out of the scope of this work. Later on, we will compare this case to case \#2, in which, the modes saturate at a higher plasma current (14 MA).

\medskip

The mode activity causes a slow flattening of the current profile as it can be observed in the evolution of the internal inductance ($l_i(3)$) in figure \ref{fig:traces_3D_1}. The MHD modes lead to the destruction of a significant fraction of the magnetic flux surfaces as shown in the Poincare plots of figure \ref{fig:poinc_1}. Almost complete ergodization of the field line topology is found at $t=13$ ms while at $t=23$ ms flux-surfaces reappear at the plasma core. The confinement of runaway electrons in the 3D perturbed fields produced with this simulation is studied in \cite{konstaREVDE}, showing that they will be quickly depleted at $t=13$ ms but that they could re-appear in the re-formed core flux surfaces ($t=23$ ms). 

\medskip

As $I_p$ decays, the plasma moves vertically upwards and toroidal current in the halo region is induced (see $I_p$, $Z_{axis}$ and $I_{halo,\phi}$ traces in figure \ref{fig:traces_3D_1}). The CQ time defined as $\tau_{CQ}\equiv(t_{20\%}-t_{80\%})/0.6$ \cite{eidietis2015itpa} is 47 ms,  which is close to the envisaged minimum CQ time for mitigated disruptions (50 ms) \cite{Sugihara2007_forces}.   Such evolution leads to an edge safety factor ($q_{95}$ trace) that remains in a range of 3-4. The effects of the halo currents in the evolution of $q_{95}$ will be discussed in section \ref{sec:halos}.

\medskip

Since $q_{95}$ stays above the value of 3, modes that could result in large horizontal wall forces are not observed (i.e. $m/n=1/1$ modes). Accordingly, the horizontal force  remains at the noise level (see $\sqrt{F_x^2+F_y^2}$ trace in figure \ref{fig:traces_3D_1}). The maximum toroidal peaking factor of the poloidal halo currents (TPF trace in figure \ref{fig:traces_3D_1}) is 1.2 and arises when the halo current fraction (HF) is still very small ($2\%$). 
 We define the halo fraction as the total poloidal halo current ($I_{halo,\theta}$) normalized by the pre-disruptive toroial plasma current. The TPF is defined as
\begin{equation}
\textrm{TPF} \equiv \max\left( \frac{1}{2} \oint_\phi |\v{J}\cdot\v{n}|R dl \right) / \left(I_{halo,\theta}/2\pi\right)
\end{equation}
where $\v{J}\cdot\v{n}$ is the normal current density into the wall and $dl$ is the differential poloidal length of the wall's contour at a given $\phi$. The maximum value of the HF$\times$TPF product is 0.09, which is significantly lower than the largest products of 0.75 observed in current experiments \cite{eidietis2015itpa} as it is expected for mitigated disruptions. A value of the HF$\times$TPF product below 0.15 corresponds to category I (frequent occurrence) electromagnetic transients considered for the ITER design \cite{lehnen2015disruptions}.

\begin{figure}[h]
  \centering
  \includegraphics[width=0.45\textwidth]{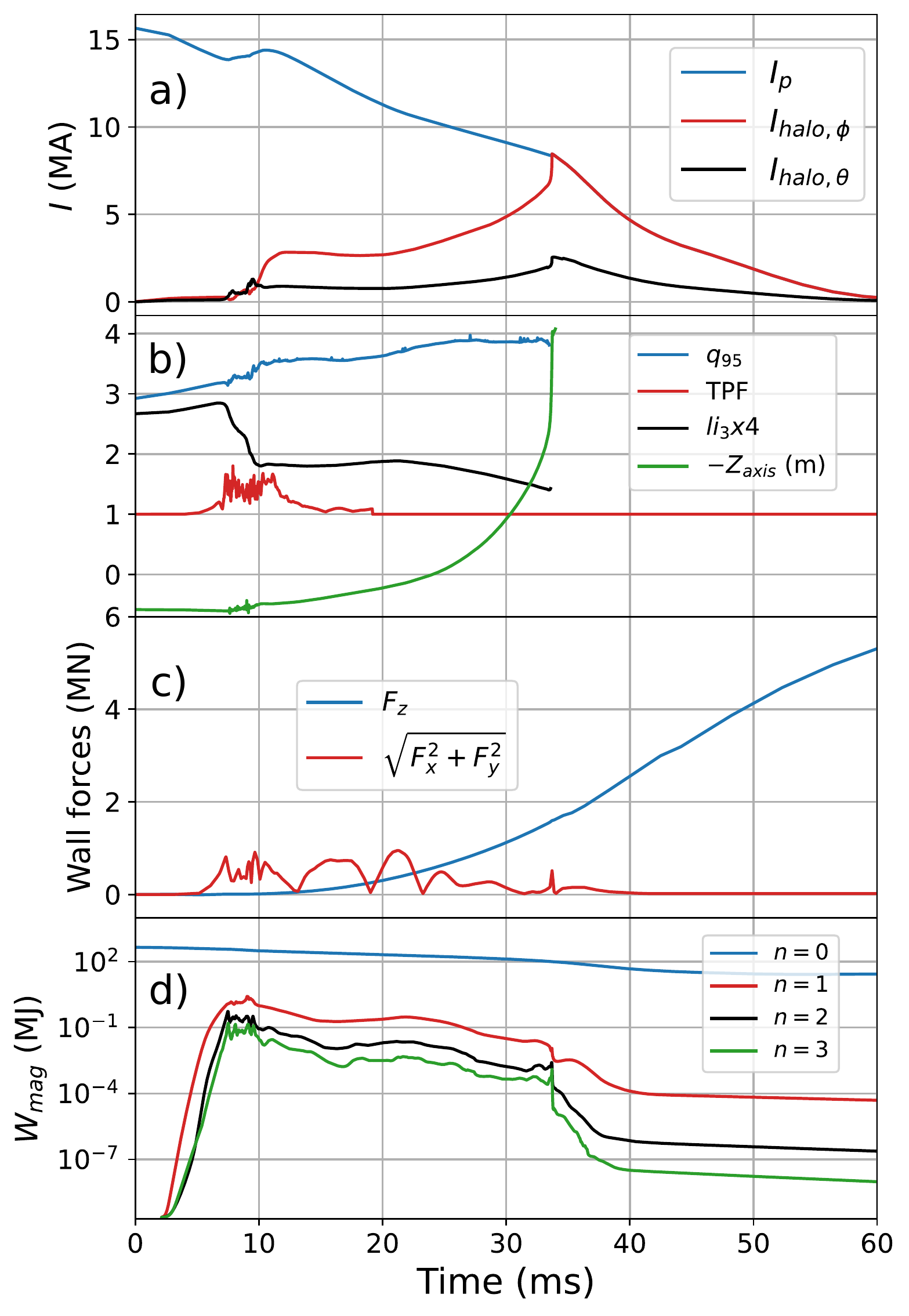}
\caption{Time traces of the CQ simulation for case \#2. (a) Total plasma toroidal current ($I_p$), total poloidal ($I_{halo,\theta}$) and toroidal  ($I_{halo,\phi}$)  halo currents . (b) Edge safety factor ($q_{95}$), toroidal peaking factor of the poloidal halo currents (TPF), internal inductance ($l_i(3)$) and vertical position of the magnetic axis ($Z_{axis}$). (c) Vertical ($F_z$) and horizontal wall forces  ($\sqrt{F_x^2+F_y^2}$). (d) The poloidal magnetic energy of the most dominant toroidal mode numbers ($n=0,1,2,3$).  }
\label{fig:traces_3D_2}
\end{figure}

\subsection{CQ phase of case \#2}
For case \#2, the axisymmetric run lasted about 2.1 ms to establish the profiles shown in figure \ref{fig:eq_profiles} and the non-axisymmetric modes were initialized to noise level at that point in time. Similar to case \#1,  several tearing modes are unstable as shown in figure \ref{fig:modes} but here a stronger dominance of the 2/1 mode is observed. The evolution of the magnetic energy for a selection of  dominant mode numbers ($n=1,2,3$) is shown in figure \ref{fig:traces_3D_2}. 

\medskip

\begin{figure}[h]
  \centering
  \includegraphics[width=0.45\textwidth]{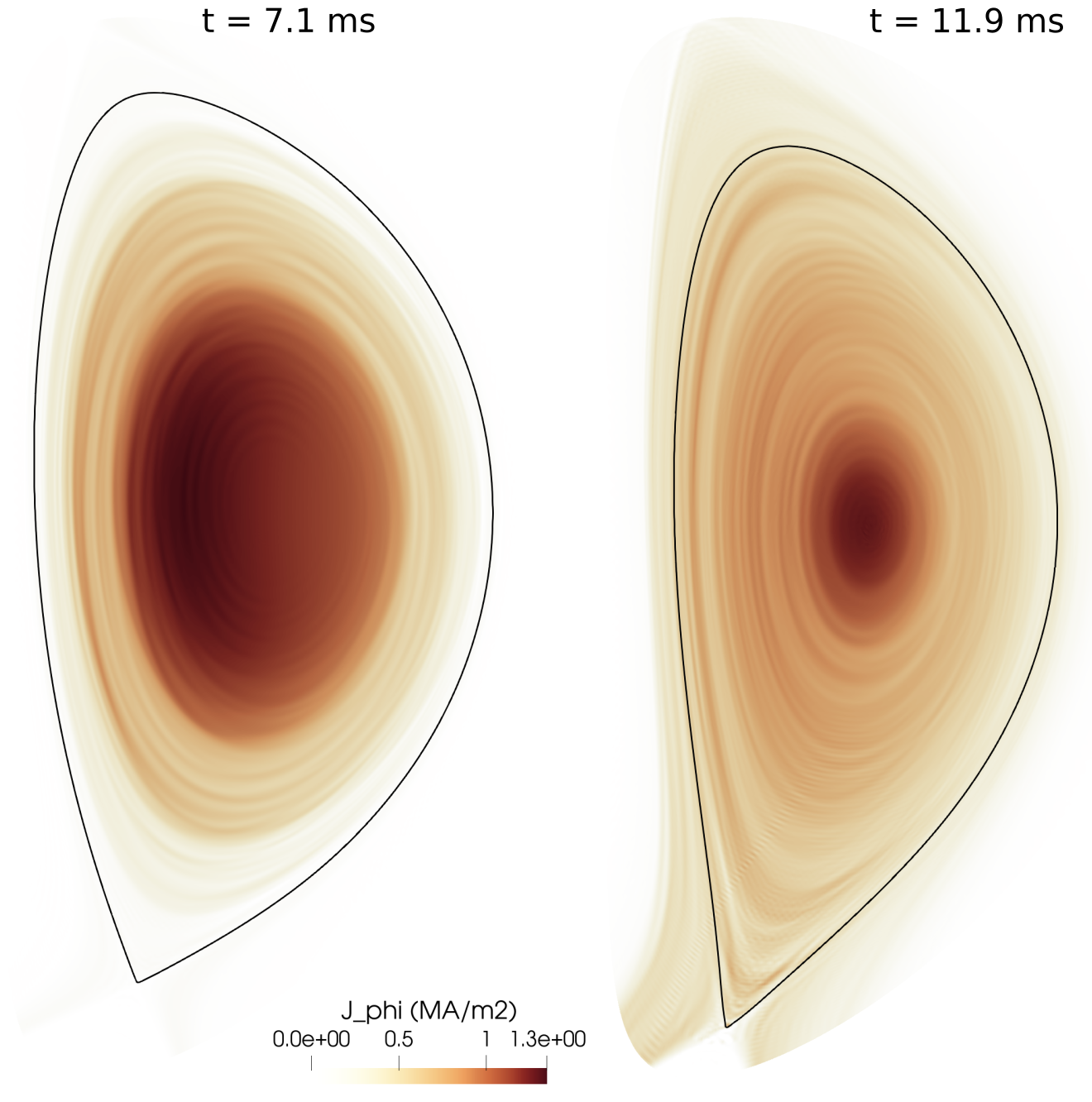}
\caption{2D distribution of the toroidal current density of the $n=0$ mode for case \# 2 at $t=7.1$ ms and $t=11.9$ ms. The black contour corresponds to the LCFS. The figure shows how the current density profile flattens beyond the LCFS.}
\label{fig:flattening2_curr}
\end{figure}

\begin{figure}[h!]
  \centering
  \includegraphics[width=0.45\textwidth]{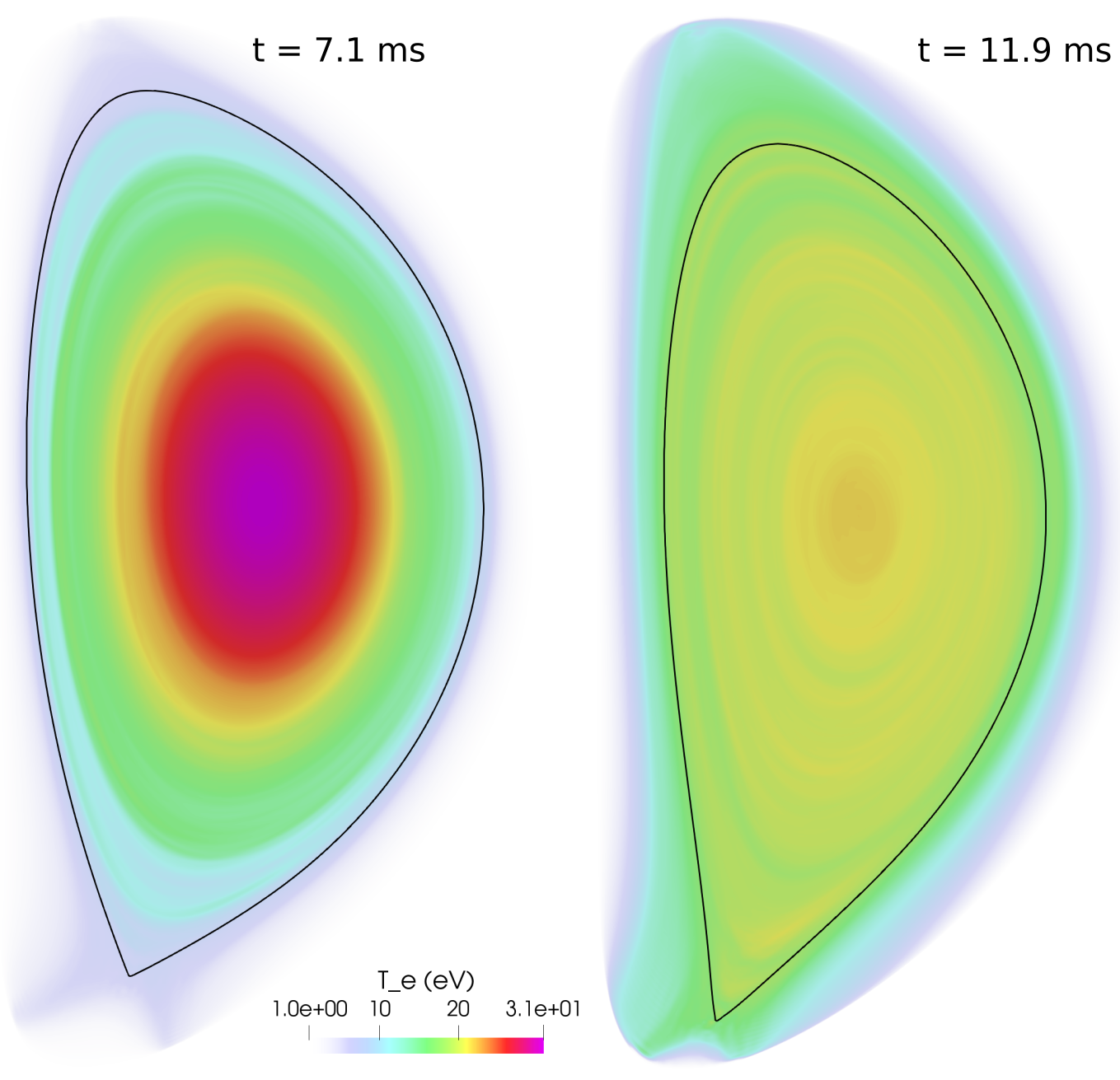}
\caption{2D distribution of the electron temperature of the $n=0$ mode for case \# 2 at $t=7.1$ ms and $t=11.9$ ms. The black contour corresponds to the LCFS. The figure shows how the temperature profile flattens beyond the LCFS.}
\label{fig:flattening2_Te}
\end{figure}

Contrary to case \#1, the saturation of the modes energy takes place at a larger fraction of the initial plasma current ($I_p\approx 14$ MA at $t=7$ ms). The non-linear growth of non-axisymmetric modes causes a large and quick flattening of the current profile as indicated by the $l_i(3)$ trace in figure \ref{fig:traces_3D_2}. It is noteworthy that the large drop in inductance ($\Delta l_i(3)=0.26$ in 3.3 ms) is followed by an increase of the current in the halo region and a spike in the total current. This suggests that the current flattening caused by the MHD activity takes place also beyond the LCFS, and that scrape-off layer currents may be playing a important role when describing the typical $I_p$-spikes that are routinely observed in disruptive plasmas. The distributions of the current and temperature before and after the flattening caused by the MHD activity are presented in figures \ref{fig:flattening2_curr} and \ref{fig:flattening2_Te} respectively.

\medskip

The resulting current quench time for this simulation is $\tau_{CQ}=45.5$ ms, which is also close to the minimum allowed by ITER specifications (50 ms). In contrast with case \#1, the plasma drifts vertically downwards as $I_p$ decays (note the minus sign in front of the $Z_{axis}$ trace in figure \ref{fig:traces_3D_2}). We attribute this change of direction to the stronger drop in $l_i(3)$ for case \#2. It was found in \cite{lukash2005analysis} that large drops of the internal inductance during major disruptions in ITER lead to downward displacements. Similarly for the ASDEX-Upgrade tokamak \cite{nakamura1996acceleration}, it was found that drops of $l_i$ in lower single-null plasmas cause a "vertical dragging effect" that moves the plasma downwards.

\medskip

Similar to case \#1, $q_{95}$ remains in the range of $3-4$ during the CQ and the horizontal force remains at noise level (below 1 MN). The maximum TPF is 1.7 and arises when the halo fraction is 5.7\%. The maximum HF$\times$TPF product is 0.17 which is two times larger than in case \#1. This is still far from the largest experimental products (0.75), but slightly above the level for category I electromagnetic transients in ITER (0.15).

\begin{figure}[h]
  \centering
  \includegraphics[width=0.45\textwidth]{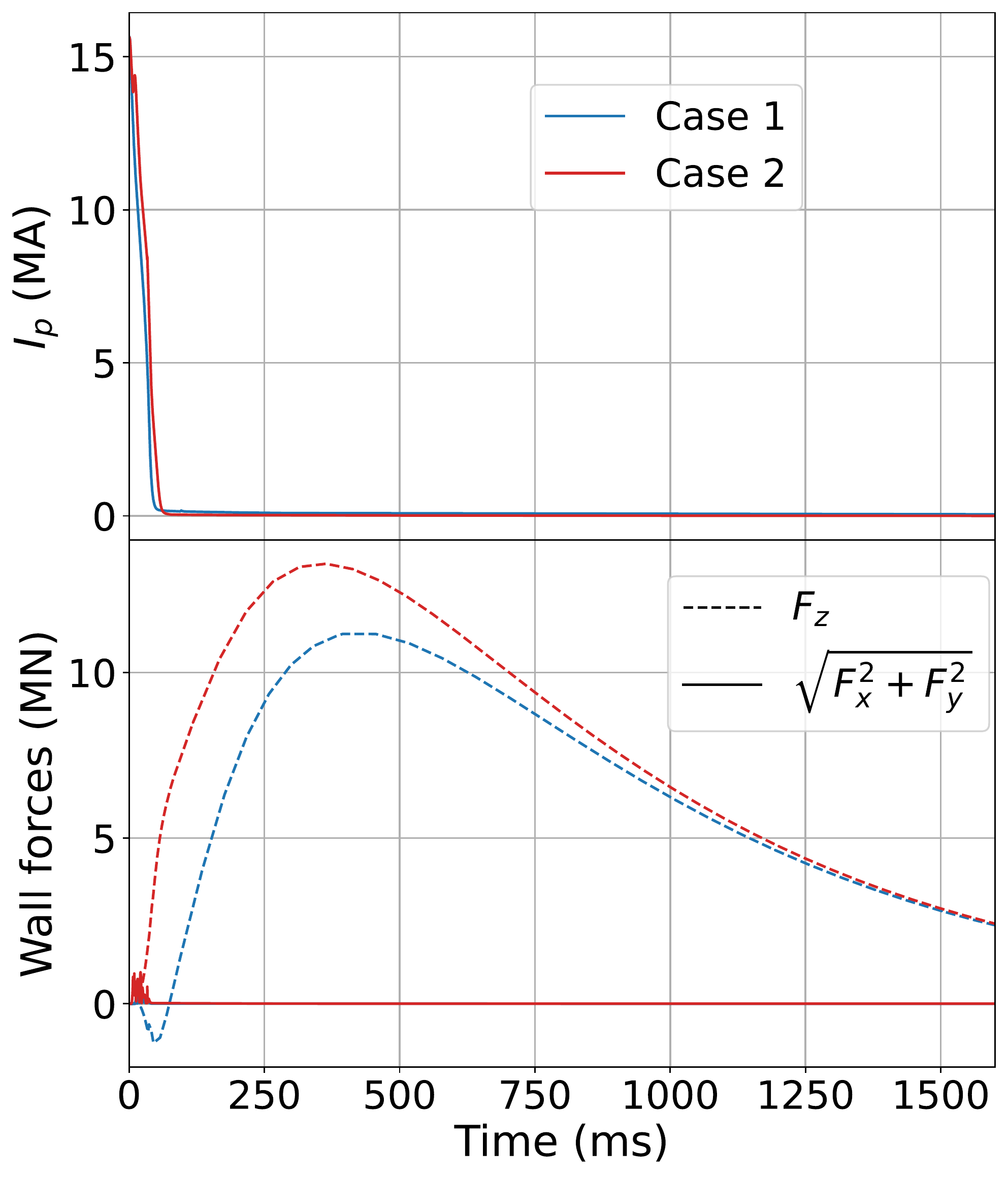}
\caption{Total plasma current and wall forces for the simulated cases. }
\label{fig:wallforces}
\end{figure}

\subsection{Wall forces after the current quench}

\begin{figure}[h]
  \centering
  \includegraphics[width=0.25\textwidth]{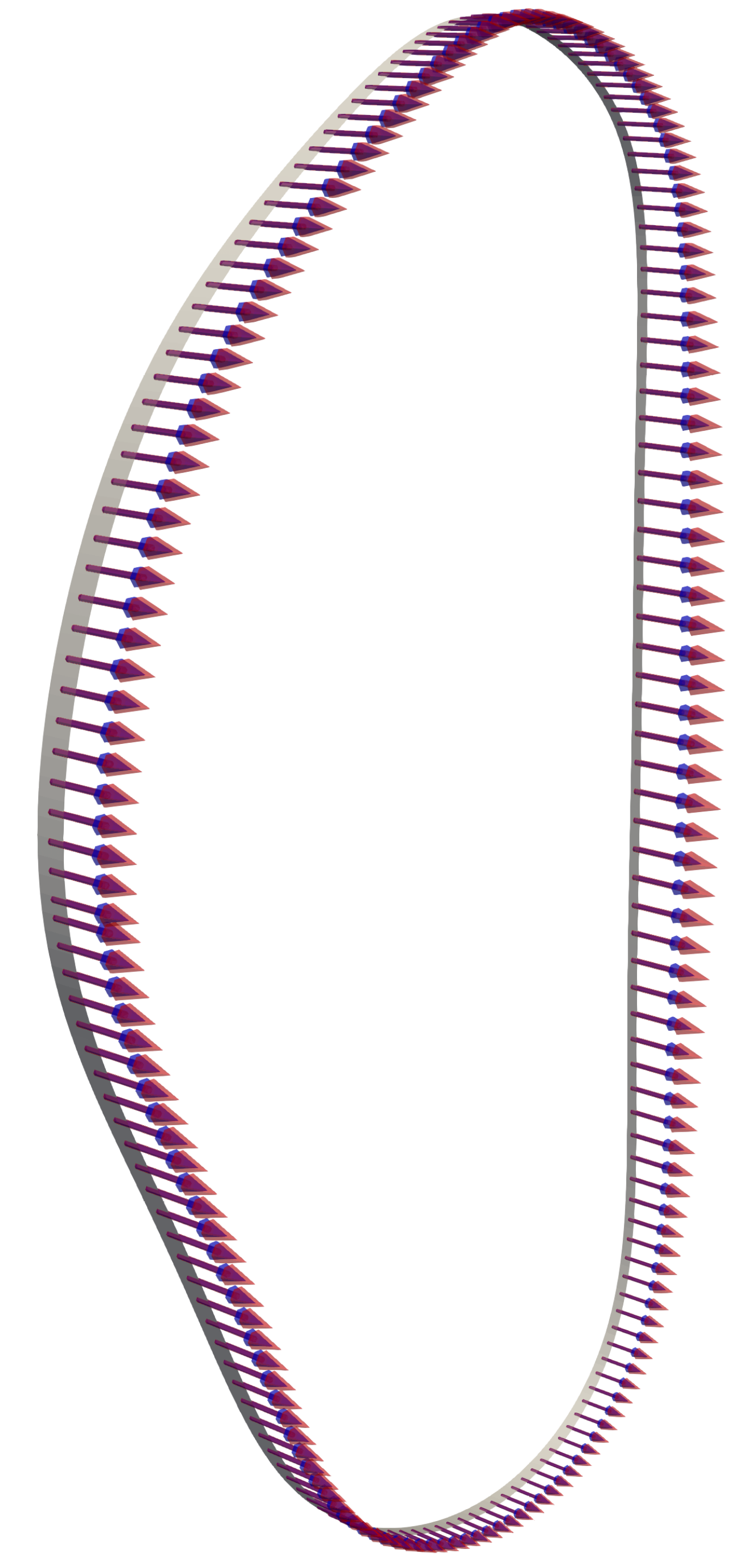}
\caption{Current density distribution in the inner vessel at $\phi=0$ when the vertical force is maximum. The blue and red arrows represent the current density magnitude of case \#1 and case \#2 respectively. }
\label{fig:wallcurr}
\end{figure}

We continue the simulation after the plasma current has completely decayed. During that phase, induced wall currents during the CQ relax and decay, leading to the penetration of the magnetic field across the wall and to the rise of net wall forces. As seen in figure \ref{fig:wallforces}, the vertical force arises in a time scale given by the $L/R$ time of the vacuum vessel ($\sim$ 500 ms). Note that the time traces of the vertical forces are very similar to the ones obtained in \cite{albanese2015effects} for ITER disruptions with volumetric walls.  Both cases show similar vertical maximal forces in the range of 11-14 MN despite the different directions for the vertical displacement. At the time of maximal vertical force, the distribution of the current density in the inner vacuum vessel is roughly identical (see figure \ref{fig:wallcurr}). In addition, we checked that the current distribution in the other passive components are also similar for both cases. Since the wall force and the final current distribution is independent of the vertical motion direction, we conclude that the vertical force arises due to the diffusion of the net toroidal wall current along the wall contour and its interaction with the magnetic field produced by the poloidal field coils. Such force would in principle not arise in the case of a fully up-down symmetric plasma centered in an up-down symmetric wall before the CQ starts, which points to the advantage of having up/down symmetric plasmas and reactor designs to minimize disruption forces. However as depicted in figure \ref{fig:grid_wall}, the plasma and wall structures are not up-down symmetric in ITER and therefore, it is natural that these asymmetries lead to net wall forces due to a net toroidal current. In any case, the magnitude of the maximum $F_z$ is a factor 6 smaller than the largest estimates \cite{Sugihara2007_forces}. Finally the horizontal force completely decays at the moment the plasma current vanishes, confirming that the found horizontal force is noise arising from a finite employed resolution.

\subsection{Influence of halo currents and other parameters}
\label{sec:halos}
In our model the temperature at the plasma-wall interface is initialized with an electron temperature of 1 eV. As the plasma is scraped-off due to the vertical motion and as the ergodic field lines connect the plasma core with the halo region, the SOL temperature increases. As shown in figure \ref{fig:flattening2_Te}, the temperature can reach values of 10-15 eV in this region during the CQ. Such temperatures are consistent with experimental temperature measurements at the plasma-wall interface during disruptions \cite{Artola_2021}. At these temperatures, the Spitzer-Harm parallel conductivity ($\kappa_\parallel\propto T_e^{2.5}$)  remains moderate and allows significant temperatures / temperature gradients in the SOL to carry the corresponding heat-flux. Estimates for ITER field line lengths of $100$ m at $T_e=10$ eV and $n_e=10^{20}$ m$^{-3}$ give characteristic parallel conduction times of the order of 70 ms.

\medskip

Parallel convection was found not to be sufficient to decrease the SOL temperature to lower values, although we could not run the case at lower parallel viscosities due to the appearance of numerical instabilities in the $v_\parallel$ variable. In addition, the parallel viscosity had to be increased by an extra factor of 10 at some points of the simulations to deal with such numerical problems (specifically at $t=14.3$ ms for case \#1 and at $t=19.8$ ms for case \#2). However, when comparing re-run fragments of the simulation with and without the increase of $\mu_\parallel$ we did not observe a significant difference in the evolution of $q_{95}$, the magnetic energies or the wall forces. Note here that convection can transport thermal energy towards the plasma-wall interface, but this energy can only be lost there by conduction since we set a no-normal flow condition ($\v{v}\cdot\v{n}=0$).

\medskip

We note that stronger energy sinks in the halo region could decrease the temperature and consequently the total halo currents. In principle, the total halo currents can change the evolution of $q_{95}$. When the halo current flow is limited to very low values due to very low temperatures in the halo, the current that otherwise would be induced in the halo can be partially re-induced in the core, thus slowing down the $I_p$ decay with respect to the vertical motion and leading to smaller $q_{95}$ values that can potentially drive larger asymmetries in the halo currents. The maximal toroidal halo currents found here (4.1 MA for case \#1 and 8.4 MA for case \#2) and the maximal poloidal halo current fractions (8.7\% for case \#1 and 16.6\% for case \#2) are within (or very close for case \#2) to the ITER specifications for category I electromagnetic transients.


\medskip

More advanced SOL models including Ohmic heating, impurity transport and radiation, neutral particles and sheath boundary conditions are needed to reliably predict the temperatures and densities in the halo region and thus the correct evolution of $q_{95}$. In future works we will pursue such simulations. 

\section{Conclusions}
\label{sec:conclusions}
We have presented complete 3D MHD simulations of the current quench phase of ITER mitigated plasmas in a comprehensive manner. The chosen parameters include realistic Lundquist numbers (or Spitzer resistivities) and parallel heat conductivity.  Due to the high computational cost of such simulations (of the order of 5 million core.hours per simulation) we present two cases featuring an upwards and a downwards vertical displacement. Additional scans will be performed in further follow-up studies. 

\medskip

Both cases show that the integral wall forces will be highly reduced for mitigated plasmas with respect to the largest expected values. We find horizontal forces below 1 MN, with the largest extrapolations from JET being $\sim$ 40 MN. We attribute the absence of the horizontal forces to the lack of 1/1 modes as  $q_{95}$ remains above the value of 3 during the current quench. Such beneficial evolution of $q_{95}$ originates from the fast current quench time ($\sim 50$ ms) compared to the wall current decay time ($\sim 500$ ms) and the induced toroidal halo currents. Further analysis including more advanced models of the halo region will be pursued in future studies. Apart from more refined models of the SOL, we will also pursue studies including the toroidal asymmetries of the ITER vacuum vessel that have not been taken into account in this work. The vessel asymmetries need further assessment since they can potentially impact the current flow and thus the asymmetries of forces during ITER disruptions.  

\medskip

A maximum vertical force of 11-14 MN is found regardless of the direction of the plasma vertical motion. Such force is attributed to the diffusion of the net wall current in an up-down asymmetric vessel and its interaction with the poloidal field coils. The simulated mitigated disruptions show a reduction of the vertical force by a factor of 6 with respect to the largest expected values (80 MN), which confirms the effectiveness of the ITER mitigation strategy and that halo currents are within the category I boundary. For disruptions with CQ times much shorter than the wall's resistive time, our simulations also point to the fact that disruption forces can be minimized if the pre-TQ plasma and vessel structures are up/down symmetric,  which has implications for DEMO reactor designs. 

\medskip

We also highlight the potential importance of considering the SOL to describe the $I_p$-spike phenomenon that is typically observed in tokamak disruptions. As it has been shown in this work, the flattening of the toroidal current and temperature profiles due to the MHD activity takes place beyond the Last Closed Flux Surface.

\medskip

Finally we conclude that the horizontal wall force rotation is not a concern for these highly mitigated plasma disruptions in ITER, since when the wall forces form, there are no remaining sources of rotation (i.e. the plasma current has decayed already).

\section*{Acknowledgements}
ITER is the Nuclear Facility INB no. 174. This paper explores physics processes during the plasma operation of the tokamak when disruptions take place; nevertheless the nuclear operator is not constrained by the results presented here. The views and opinions expressed herein do not necessarily reflect those of the ITER Organization. The simulations presented here have been performed using the ITER HPC cluster, the Marconi-Fusion supercomputer and Google Cloud resources provided to the ITER Organization under the Cloud computing proof of concept collaboration. We would like to thank S. Pinches, Frederic Hamiez and Peter Kroul for their technical assistance  with these simulations. This work has been carried out within the framework of the EUROfusion Consortium and has received funding from the Euratom research and training program 2014-2018 and 2019-2020 under grant agreement No 633053. The views and opinions expressed herein do not necessarily reflect those of the European Commission.


\bibliography{biblio.bib}{}
\bibliographystyle{abbrvurl}

\end{document}

%% file: packages.tex
\usepackage[utf8]{inputenc}
\usepackage[english]{babel}
\usepackage[colorlinks,linkcolor=blue,linktoc=page, citecolor=green, urlcolor=blue]{hyperref}
\usepackage{makeidx}
\usepackage[T1]{fontenc}
\usepackage{mathrsfs}
\usepackage{dsfont}

\usepackage{lmodern}
\usepackage[margin=0.8in]{geometry}
\usepackage[titletoc]{appendix}
\usepackage{amsfonts}
\usepackage{graphicx}
\usepackage{tcolorbox}
\tcbuselibrary{theorems}
\usepackage{lipsum}
\usepackage{marvosym} 
\usepackage{appendix}
\usepackage{ntheorem}
\usepackage{pbsi}
\usepackage[printonlyused]{acronym}
\usepackage{cancel}
\usepackage{empheq}

\usepackage{tikz}
\usetikzlibrary{positioning,arrows}
\usetikzlibrary{decorations.pathmorphing}
\usetikzlibrary{decorations.markings}

\usepackage{tabulary}
\usepackage{slashed}
\usepackage{esvect}
\usepackage{amsmath}
\usepackage{mathtools}

\usepackage{nccmath}
\usepackage{pifont}
\usepackage{colortbl}
\usepackage{xcolor}
\usepackage{cancel}
\usepackage{simplewick}
\usepackage{wrapfig}
\usepackage[font=small,labelfont=sc]{caption}
\usepackage{subcaption}

\renewcommand{\v}[1]{\ensuremath{\mathbf{#1}}} 
 
\let\baraccent=\= 
\renewcommand{\=}[1]{\stackrel{#1}{=}} 

%% file: ITER_wall_forces.bbl
\begin{thebibliography}{10}

\bibitem{albanese2015effects}
R.~Albanese, B.~Carpentieri, M.~Cavinato, S.~Minucci, R.~Palmaccio, A.~Portone,
  G.~Rubinacci, P.~Testoni, S.~Ventre, and F.~Villone.
\newblock Effects of asymmetric vertical disruptions on {ITER} components.
\newblock {\em Fusion Engineering and Design}, 94:7--21, 2015.
\newblock \href {https://doi.org/10.1016/j.fusengdes.2015.02.034}
  {\path{doi:10.1016/j.fusengdes.2015.02.034}}.

\bibitem{artola2020understanding}
F.~Artola, K.~Lackner, G.~Huijsmans, M.~Hoelzl, E.~Nardon, and A.~Loarte.
\newblock Understanding the reduction of the edge safety factor during hot
  {VDE}s and fast edge cooling events.
\newblock {\em Physics of Plasmas}, 27(3):032501, 2020.
\newblock \href {https://doi.org/10.1063/1.5140230}
  {\path{doi:10.1063/1.5140230}}.

\bibitem{artola2018non}
F.~J. Artola, G.~Huijsmans, M.~Hoelzl, P.~Beyer, A.~Loarte, and Y.~Gribov.
\newblock Non-linear magnetohydrodynamic simulations of edge localised mode
  triggering via vertical position oscillations in {ITER}.
\newblock {\em Nuclear Fusion}, 2018.
\newblock \href {https://doi.org/10.1088/1741-4326/aace0e}
  {\path{doi:10.1088/1741-4326/aace0e}}.

\bibitem{Artola_2021}
F.~J. Artola, A.~Loarte, E.~Matveeva, J.~Havlicek, T.~Markovic, J.~Adamek,
  J.~Cavalier, L.~Kripner, G.~Huijsmans, M.~Lehnen, M.~Hoelzl, and R.~Panek.
\newblock Simulations of {COMPASS} vertical displacement events with a
  self-consistent model for halo currents including neutrals and sheath
  boundary conditions.
\newblock {\em Plasma Physics and Controlled Fusion}, apr 2021.
\newblock \href {https://doi.org/10.1088/1361-6587/abf620}
  {\path{doi:10.1088/1361-6587/abf620}}.

\bibitem{artola3DVDE_bench}
F.~J. Artola, C.~R. Sovinec, S.~C. Jardin, M.~Hoelzl, I.~Krebs, and C.~Clauser.
\newblock 3d simulations of vertical displacement events in tokamaks: {A}
  benchmark of {M3D-C1}, {NIMROD}, and {JOREK}.
\newblock {\em Physics of Plasmas}, 28(5):052511, 2021.
\newblock \href {https://doi.org/10.1063/5.0037115}
  {\path{doi:10.1063/5.0037115}}.

\bibitem{artolasuch:tel-02012234}
F.~Artola~Such.
\newblock {\em Free-boundary simulations of {{{MHD}}} plasma instabilities in
  tokamaks}.
\newblock Theses, {Universit{\'e} Aix Marseille}, Nov. 2018.
\newblock URL: \url{https://tel.archives-ouvertes.fr/tel-02012234}.

\bibitem{bachmann2011specification}
C.~Bachmann, M.~Sugihara, R.~Roccella, G.~Sannazzaro, Y.~Gribov, V.~Riccardo,
  T.~Hender, S.~Gerasimov, G.~Pautasso, A.~Belov, et~al.
\newblock Specification of asymmetric vde loads of the {ITER} tokamak.
\newblock {\em Fusion Engineering and Design}, 86(9-11):1915--1919, 2011.
\newblock \href {https://doi.org/10.1016/j.fusengdes.2011.02.096}
  {\path{doi:10.1016/j.fusengdes.2011.02.096}}.

\bibitem{Clauser_2019}
C.~Clauser, S.~Jardin, and N.~Ferraro.
\newblock Vertical forces during vertical displacement events in an {{ITER}}
  plasma and the role of halo currents.
\newblock {\em Nuclear Fusion}, 59(12):126037, oct 2019.
\newblock URL: \url{https://doi.org/10.1088%2F1741-4326%2Fab440a}, \href
  {https://doi.org/10.1088/1741-4326/ab440a}
  {\path{doi:10.1088/1741-4326/ab440a}}.

\bibitem{eidietis2015itpa}
N.~Eidietis, S.~Gerhardt, R.~Granetz, Y.~Kawano, M.~Lehnen, J.~Lister,
  G.~Pautasso, V.~Riccardo, R.~Tanna, A.~Thornton, et~al.
\newblock The {ITPA} disruption database.
\newblock {\em Nuclear Fusion}, 55(6):063030, 2015.
\newblock \href {https://doi.org/10.1088/0029-5515/55/6/063030}
  {\path{doi:10.1088/0029-5515/55/6/063030}}.

\bibitem{gerasimov2015jet}
S.~Gerasimov, P.~Abreu, M.~Baruzzo, V.~Drozdov, A.~Dvornova, J.~Havlicek,
  T.~Hender, O.~Hronova, U.~Kruezi, X.~Li, et~al.
\newblock {JET} and {COMPASS} asymmetrical disruptions.
\newblock {\em Nuclear Fusion}, 55(11):113006, 2015.
\newblock \href {https://doi.org/10.1088/0029-5515/55/11/113006}
  {\path{doi:10.1088/0029-5515/55/11/113006}}.

\bibitem{overview}
M.~Hoelzl, G.~Huijsmans, S.~Pamela, M.~Becoulet, E.~Nardon, F.~Artola, and
  B.~N. et~al.
\newblock The {JOREK} non-linear extended {{{MHD}}} code and applications to
  large-scale instabilities and their control in magnetically confined fusion
  plasmas.
\newblock {\em Nuclear Fusion (submitted)}, 2020.
\newblock URL: \url{https://arxiv.org/abs/2011.09120}.

\bibitem{Hoelzl_2021}
M.~Hoelzl, G.~Huijsmans, S.~Pamela, M.~B{\'{e}}coulet, E.~Nardon, F.~Artola,
  B.~Nkonga, C.~Atanasiu, V.~Bandaru, A.~Bhole, D.~Bonfiglio, A.~Cathey,
  O.~Czarny, A.~Dvornova, T.~Feh{\'{e}}r, A.~Fil, E.~Franck, S.~Futatani,
  M.~Gruca, H.~Guillard, J.~Haverkort, I.~Holod, D.~Hu, S.~Kim, S.~Korving,
  L.~Kos, I.~Krebs, L.~Kripner, G.~Latu, F.~Liu, P.~Merkel, D.~Meshcheriakov,
  V.~Mitterauer, S.~Mochalskyy, J.~Morales, R.~Nies, N.~Nikulsin, F.~Orain,
  J.~Pratt, R.~Ramasamy, P.~Ramet, C.~Reux, K.~Särkimäki, N.~Schwarz, P.~S.
  Verma, S.~Smith, C.~Sommariva, E.~Strumberger, D.~van Vugt, M.~Verbeek,
  E.~Westerhof, F.~Wieschollek, and J.~Zielinski.
\newblock The {JOREK} non-linear extended {{{MHD}}} code and applications to
  large-scale instabilities and their control in magnetically confined fusion
  plasmas.
\newblock {\em Nuclear Fusion}, 61(6):065001, May 2021.
\newblock \href {https://doi.org/10.1088/1741-4326/abf99f}
  {\path{doi:10.1088/1741-4326/abf99f}}.

\bibitem{Hoelzl2012a}
M.~Hoelzl, P.~Merkel, G.~T.~A. Huysmans, E.~Nardon, E.~Strumberger, R.~McAdams,
  I.~Chapman, S.~Günter, and K.~Lackner.
\newblock Coupling {JOREK} and {STARWALL} codes for non-linear resistive-wall
  simulations.
\newblock {\em Journal of Physics: Conference Series}, 401:012010, dec 2012.
\newblock \href {https://doi.org/10.1088/1742-6596/401/1/012010}
  {\path{doi:10.1088/1742-6596/401/1/012010}}.

\bibitem{huysmans2007mhd}
G.~Huysmans and O.~Czarny.
\newblock {{{MHD}}} stability in {X}-point geometry: simulation of {ELM}s.
\newblock {\em Nuclear fusion}, 47(7):659, 2007.
\newblock \href {https://doi.org/10.1088/0029-5515/47/7/016}
  {\path{doi:10.1088/0029-5515/47/7/016}}.

\bibitem{Kiramov_2018}
D.~I. Kiramov and B.~N. Breizman.
\newblock Force-free motion of a cold plasma during the current quench.
\newblock {\em Physics of Plasmas}, 25(9):092501, sep 2018.
\newblock \href {https://doi.org/https://doi.org/10.1063/1.5046517}
  {\path{doi:https://doi.org/10.1063/1.5046517}}.

\bibitem{krebs2020axisymmetric}
I.~Krebs, F.~Artola, C.~Sovinec, S.~Jardin, K.~Bunkers, M.~Hoelzl, and
  N.~Ferraro.
\newblock Axisymmetric simulations of vertical displacement events in tokamaks:
  A benchmark of {M3D-C1}, {NIMROD}, and {JOREK}.
\newblock {\em Physics of Plasmas}, 27(2):022505, 2020.
\newblock \href {https://doi.org/10.1063/1.5127664}
  {\path{doi:10.1063/1.5127664}}.

\bibitem{lehnen2015disruptions}
M.~Lehnen, K.~Aleynikova, P.~Aleynikov, D.~Campbell, P.~Drewelow, N.~Eidietis,
  Y.~Gasparyan, R.~Granetz, Y.~Gribov, N.~Hartmann, et~al.
\newblock Disruptions in {{ITER}} and strategies for their control and
  mitigation.
\newblock {\em Journal of Nuclear Materials}, 463:39--48, 2015.
\newblock \href {https://doi.org/10.1016/j.jnucmat.2014.10.075}
  {\path{doi:10.1016/j.jnucmat.2014.10.075}}.

\bibitem{lukash2005analysis}
V.~Lukash, M.~Sugihara, Y.~Gribov, and H.~Fujieda.
\newblock Analysis of the direction of plasma vertical movement during major
  disruptions in iter.
\newblock {\em Plasma physics and controlled fusion}, 47(12):2077, 2005.
\newblock \href {https://doi.org/10.1088/0741-3335/47/12/001}
  {\path{doi:10.1088/0741-3335/47/12/001}}.

\bibitem{Merkel2015}
P.~Merkel and E.~Strumberger.
\newblock Linear {{{MHD}}} stability studies with the {STARWALL} code.
\newblock {\em arXiv e-prints}, arXiv:1508.04911, 2015.
\newblock URL: \url{http://arxiv.org/abs/1508.04911}.

\bibitem{nakamura1996acceleration}
Y.~Nakamura, R.~Yoshino, N.~Pomphrey, and S.~C. Jardin.
\newblock Acceleration mechanism of vertical displacement event and its
  amelioration in tokamak disruptions.
\newblock {\em Journal of nuclear science and technology}, 33(8):609--619,
  1996.
\newblock \href {https://doi.org/10.1080/18811248.1996.9731967}
  {\path{doi:10.1080/18811248.1996.9731967}}.

\bibitem{Nardon2021}
E.~Nardon, D.~Hu, F.~J. Artola, D.~Bonfiglio, M.~Hoelzl, A.~Boboc, P.~Carvalho,
  S.~Gerasimov, G.~Huijsmans, V.~Mitterauer, N.~Schwarz, and H.~Sun.
\newblock Thermal quench and current profile relaxation dynamics in
  massive-material-injection-triggered tokamak disruptions.
\newblock 63(11):115006, sep 2021.
\newblock \href {https://doi.org/10.1088/1361-6587/ac234b}
  {\path{doi:10.1088/1361-6587/ac234b}}.

\bibitem{noll1997present}
P.~Noll, P.~Andrew, M.~Buzio, R.~Litunovsky, T.~Raimondi, V.~Riccardo, and
  M.~Verrecchia.
\newblock Present understanding of electromagnetic behaviour during disruptions
  in {JET}.
\newblock In {\em Fusion Technology 1996}, pages 751--754. Elsevier, 1997.
\newblock \href {https://doi.org/10.1016/b978-0-444-82762-3.50157-9}
  {\path{doi:10.1016/b978-0-444-82762-3.50157-9}}.

\bibitem{Pautasson2016_reduction}
G.~Pautasso, M.~Bernert, M.~Dibon, B.~Duval, R.~Dux, E.~Fable, J.~C. Fuchs,
  G.~D. Conway, L.~Giannone, A.~Gude, A.~Herrmann, M.~Hoelzl, P.~J. McCarthy,
  A.~Mlynek, M.~Maraschek, E.~Nardon, G.~Papp, S.~Potzel, C.~Rapson,
  B.~Sieglin, W.~Suttrop, W.~Treutterer, and and.
\newblock Disruption mitigation by injection of small quantities of noble gas
  in {ASDEX} upgrade.
\newblock 59(1):014046, nov 2016.
\newblock \href {https://doi.org/10.1088/0741-3335/59/1/014046}
  {\path{doi:10.1088/0741-3335/59/1/014046}}.

\bibitem{Pustovitov_2017}
V.~Pustovitov, G.~Rubinacci, and F.~Villone.
\newblock On the computation of the disruption forces in tokamaks.
\newblock {\em Nuclear Fusion}, 57(12):126038, oct 2017.
\newblock \href {https://doi.org/10.1088/1741-4326/aa8876}
  {\path{doi:10.1088/1741-4326/aa8876}}.

\bibitem{Pusto2021}
V.~Pustovitov, G.~Rubinacci, and F.~Villone.
\newblock Sideways force due to coupled rotating kink modes in tokamaks.
\newblock 61(3):036018, feb 2021.
\newblock \href {https://doi.org/10.1088/1741-4326/abce3e}
  {\path{doi:10.1088/1741-4326/abce3e}}.

\bibitem{Riccardo_2000}
V.~Riccardo, P.~Noll, and S.~Walker.
\newblock Forces between plasma, vessel and {TF} coils during {AVDEs} at {JET}.
\newblock {\em Nuclear Fusion}, 40(10):1805--1810, oct 2000.
\newblock \href {https://doi.org/10.1088/0029-5515/40/10/311}
  {\path{doi:10.1088/0029-5515/40/10/311}}.

\bibitem{schioler2011dynamic}
T.~Schioler, C.~Bachmann, G.~Mazzone, and G.~Sannazzaro.
\newblock Dynamic response of the {{ITER}} tokamak during asymmetric {VDE}s.
\newblock {\em Fusion Engineering and Design}, 86(9-11):1963--1966, 2011.
\newblock \href {https://doi.org/10.1016/j.fusengdes.2010.11.016}
  {\path{doi:10.1016/j.fusengdes.2010.11.016}}.

\bibitem{strauss2018reduction}
H.~Strauss.
\newblock Reduction of asymmetric wall force in {{ITER}} disruptions with fast
  current quench.
\newblock {\em Physics of Plasmas}, 25(2):020702, 2018.
\newblock \href {https://doi.org/10.1063/1.5008813}
  {\path{doi:10.1063/1.5008813}}.

\bibitem{Sugihara2007_forces}
M.~Sugihara, M.~Shimada, H.~Fujieda, Y.~Gribov, K.~Ioki, Y.~Kawano,
  R.~Khayrutdinov, V.~Lukash, and J.~Ohmori.
\newblock Disruption scenarios, their mitigation and operation window in
  {{ITER}}.
\newblock 47(4):337--352, mar 2007.
\newblock \href {https://doi.org/10.1088/0029-5515/47/4/012}
  {\path{doi:10.1088/0029-5515/47/4/012}}.

\bibitem{konstaREVDE}
K.~Särkimäki, F.~Artola, and M.~Hoelzl.
\newblock Confinement of passing and trapped runaway electrons in simulation of
  {ITER} current quench (in preparation).
\newblock {\em Nuclear Fusion}, 2021.

\bibitem{Vadim2021}
V.~Yanovskiy, N.~Isernia, V.~D. Pustovitov, V.~Scalera, F.~Villone,
  J.~Hromadka, M.~Imrisek, J.~Havlicek, M.~Hron, and R.~Panek.
\newblock Global forces on the {COMPASS}-{U} wall during plasma disruptions.
\newblock jul 2021.
\newblock \href {https://doi.org/10.1088/1741-4326/ac1545}
  {\path{doi:10.1088/1741-4326/ac1545}}.

\end{thebibliography}
